# Deep Learning-Driven Nonlinear Reduced-Order Models for Predicting Wave-Structure Interaction

R. Halder*, M. Damodaran** and B.C. Khoo[#]

*National University of Singapore, Temasek Laboratories 9 Engineering Drive 1, Singapore 117575*

## Abstract

Long Short-Term Memory (LSTM) network-driven Non-Intrusive Reduced Order Model (NROM) for predicting the dynamics of a floating box on the water surface in a wavemaker basin is addressed in this study. The ground truth or actual data for these wave-structure interactions (WSI) problems, namely box displacements and hydrodynamic forces and moments acting on the box due to wave interaction corresponding to a particular wave profile, are computed using the Smoothed Particle Hydrodynamics (SPH). The dimensionality of the system is first reduced using the Discrete Empirical Interpolation Method (DEIM) and the LSTM is applied to the reduced system resulting in a DEIM-LSTM network for developing a surrogate for prediction. The network is further enhanced by incorporating the physics information into the loss function resulting in a physics-informed LSTM (LSTM-PINN) for predicting the rigid body dynamics of box motion. The performance of predictions for these networks is assessed for the two-dimensional wave basin WSI problem as a proof-of-concept demonstration.

1. Introduction

During the past two decades, several Reduced Order Model (ROM) methods outlined in Carlberg et al. [1] have been instrumental in expediting the efficient computational modeling of dynamical systems implying significant benefits for flow prediction in engineered systems. The conventional Computational Fluid Dynamics (CFD) methods offer solutions to well-posed flow problems with proper initial and boundary conditions. Recent efforts in deep learning (DL) appear to offer efficient and effective solutions easily for ill-posed problems such as inverse, regression, and classification problems in diverse fields as outlined in Schmidhuber et al. [2] have gained significant attention in a wide variety of fluid mechanics applications such as the closure model for turbulence using machine learning as in Maulik and San [3] and Duraisamy et al. [4], the combined POD and LSTM network for incompressible flows with spectral proper orthogonal decomposition (SPOD) of Wang et al. [5] and the application of a dimensionality reduction method with DL networks for learning feature dynamics from noisy data sets in Lui and Wolf

*Research Scientist; e0010762@u.nus.edu; Currently Research Fellow University of Michigan, USA

** Senior Research Scientist; tslmura@nus.edu.sg

# Professor of Mechanical Engineering and Director Temasek Laboratories; tslhead@nus.edu.sg

[6]. The Physics Informed Neural Network (PINN) introduced by Raissi et al. [7] in computational continuum mechanics offers a trade-off between the availability of the data and the accuracy of the solution and the possibilities of training DL networks with low or no data to moderate and big data for effective prediction of flow problems. The application of machine learning in fluid mechanics as outlined in Brunton et al. [8] has motivated the present study in which the LSTM and LSTM-PINN network is coupled with the DEIM algorithm as the nonlinear model order reduction method outlined in Chaturantabut and Sorensen [9] to develop a surrogate for predicting a couple of wave-structure interaction (WSI) problems. The wave structure interaction (WSI) possesses different numerical challenges because of strong nonlinearities in the dynamical system and its multi-disciplinary nature as outlined by Chakrabarti et al. [10]. Although SPH is originally developed for astrophysical applications by Monaghan [11] and Monaghan and Lattanzio [12], it has been used for modelling wave generation and WSI problems in several studies such as Wen et al. [13] and Altomare et al. [14] which uses the implementation of the SPH in the open-source platform *DualSPHysics* outlined in Dominique et al. [15] which is the platform used in the present study to generate necessary data for the proposed nonlinear reduced-order model DEIM-LSTM and LSTM-PINN algorithms and also to compare the predicted result with the ground truth (or actual data). Figure 1 shows the schematic of the wave-basin setup for predicting the nonlinear wave interaction and the unsteady motion dynamics of a freely floating box on the surface of a wave in a two-dimensional wave basin as a proof-of-concept demonstration of the proposed LSTM-DEIM/LSTM-PINN-DEIM surrogates. The freely floating box considered has a length of $l_b$ of 0.3m, a height of $h_b$ of 0.2m, and a unit width in the spanwise direction and is assumed to have a density of *500kg/m³* which will float in water which has twice its density. This problem has been simulated using a single phase SPH by Ren et al. [16] and Domínguez et al. [17]. The movement of the paddle on the left will set up wave flumes in the wave basin and the wave heights are measured relative to the datum line (black solid line at the bottom of the wave basin) which corresponds to the undisturbed water level in the wave basin.

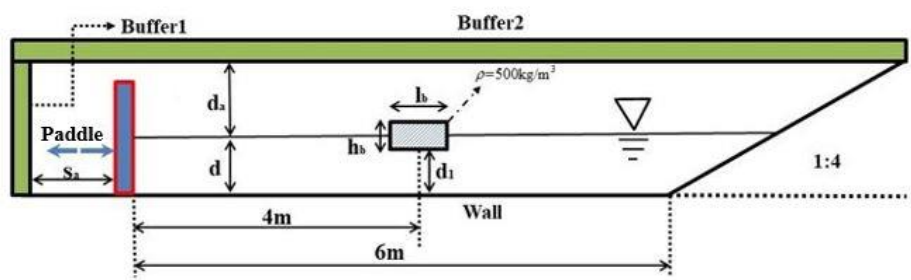

**Figure 1:** Schematic of the initial set-up and initial location for a freely floating box in the 2D wave basin.

In this study, the problem of the hydrodynamic wave interaction with a freely floating rectangular box is considered in the context of a single-phase SPH. A freely floating box is initially assumed to be still on the surface of the water. The box is excited by a wave generated from a wake maker. The wave profile has a period of $T$ of 1.2s and a wave height $H_0$ of 0.1m and its interaction with the floating box create box motions, namely surge in the direction and heave perpendicular to the direction of wave propagation and pitching rotation (pitch) about the center of mass of the box. The motion of the piston creates a second-order Stokes wave which is absorbed in the 1:4-sloped dissipative beach at the end. The no-slip boundary conditions are implemented at the bottom wall and the dissipative beach head. Buffers 1 and 2 denote the open boundaries. As this is a two-dimensional problem there will be two translational motions of the box center of mass, namely, surge and heave motions along with the $x$ and $y$ directions respectively, and a pitching motion about the center of mass. The dynamics of the floating box is defined mathematically as:

$$M \frac{dU_y}{dt} = \mathbf{F}_y + (1 - \frac{1}{\rho_r})M\mathbf{g}$$
$$M \frac{dU_x}{dt} = \mathbf{F}_x \qquad (1)$$
$$\frac{d(\mathbf{J} \cdot \boldsymbol{\omega}_s)}{dt} = \mathbf{T}$$

where $M$ is the mass of the box assumed to be 30 kg, $\mathbf{J}$ is the moment of inertia of the box about its center of mass assumed to be 0.325 kg-m$^2$, $\rho_r$ is the specific density of the box material assumed to be 0.5 and $\mathbf{F}_x$, $\mathbf{F}_y$ and $\mathbf{T}$ the forces in the surge $x$, and heave $y$ directions and torque about the box center of mass respectively. A second-order Stokes wave is generated by the wavemaker on the left end of the basin and the waves are absorbed in the 1:4-sloped dissipative beach head on the right end of the basin. No-slip boundary conditions are implemented on the bottom wall and the dissipative beach while Buffers 1 and 2 are treated as open boundaries. The total number of particles used for the single-phase SPH is 70023. A resolution of 0.01 m proposed in Ren et al. [16] is used in the present study. The box motion dynamics of Eqn. (1) in a matrix form is as follows:

$$[M]\{\ddot{X}\} = \{F\} \qquad (2.1)$$

where, $[M] = \begin{bmatrix} M & 0 & 0 \\ 0 & M & 0 \\ 0 & 0 & J \end{bmatrix}$, $\{X\} = \begin{bmatrix} S \\ h \\ \alpha \end{bmatrix}$, $\{F\} = \begin{bmatrix} F_x \\ F_y + \left(1 - \frac{1}{\rho_r}\right)M\mathbf{g} \\ \mathbf{T} \end{bmatrix}$ where $S, h$ and $\alpha$ are the surge, heave, and pitch. The first order time derivatives of these kinematic variables, i.e., $\dot{S}, \dot{h}$ and $\dot{\alpha}$ are the surge, heave and pitch rates and $\mathbf{g}$ are the acceleration due to gravity. Since $\dot{X} - \dot{X} = 0$ it follows that

$$\{\ddot{X}\} = [M^{-1}]\{F\} \Rightarrow \begin{bmatrix} \dot{X} \\ \dot{X} \end{bmatrix} + \begin{bmatrix} 0 & 0 & 0 & -1 & 0 & 0 \\ 0 & 0 & 0 & 0 & -1 & 0 \\ 0 & 0 & 0 & 0 & 0 & -1 \\ 0 & 0 & 0 & 0 & 0 & 0 \\ 0 & 0 & 0 & 0 & 0 & 0 \\ 0 & 0 & 0 & 0 & 0 & 0 \end{bmatrix} \begin{Bmatrix} X \\ \dot{X} \end{Bmatrix} = \begin{Bmatrix} 0 \\ 0 \\ 0 \\ M^{-1}F \end{Bmatrix} \quad (2.2)$$

**II. Nonlinear Reduced Order Model:**

Formulation of the nonlinear reduced order models (NROMs) proposed for the possibility of developing digital twins for the wave-structure interactions are outlined in this section. Section II(a) discusses the basic architecture of the LSTM and LSTM PINN networks and outlines the elements of the application of LSTM-PINN network for rigid body dynamics equation governing the box motion while Section II(b) addresses the coupling of the LSTM with the dimensionality reduction approach of DEIM. Although the PINN network is computationally expensive than conventional LSTM network and numerical solution of the benchmark rigid body dynamics equation, the objective of the current work is to present a proof-of-concept model for the realization of a LSTM-PINN model which can be used for predicting the response of a large-scale application using sparse or no data to assess if the potential of the LSTM-PINN network can be realized. The DEIM-LSTM and LSTM-PINN codes and data can be found out in https://github.com/rahulhalderAERO/Hydrodynamic_LSTM_ROM

**II(a) LSTM and LSTM-PINN network:**

In this current section, the formulation of functional relationship between input and output are first explained with a simpler ANN network. A single layer ANN network consisting of few neurons mapping input data, namely the hydrodynamic forces acting on a floating box computed from SPH and the output data consisting of the floating box kinematic variables computed from SPH, which form the dataset for training the surrogate for predicting the dynamics of the floating box in a wave basin. The mapping can be expressed as follows:

$$\{S \quad h \quad \alpha \quad \dot{S} \quad \dot{h} \quad \dot{\alpha}\} = f(W_2(W_1[F_x \quad F_y \quad T] + b_1) + b_2) \quad (3)$$

where $\{S \quad h \quad \alpha \quad \dot{S} \quad \dot{h} \quad \dot{\alpha}\} \in \mathbb{R}^{6 \times n_t}$ and $[F_x \quad F_y \quad T] \in \mathbb{R}^{3 \times n_t}$ are output and input data, $W_1$ and $W_2$ are the weight matrices, $b_1$ and $b_2$ are the bias vectors respectively and $f$ is the nonlinear activation function which in this study uses the *tanh* function. The structure of a simple LSTM network is different from that of the ANN architecture, and it will be demonstrated for the same input and output datasets associated with the box dynamics floating on the hydrodynamics waves. The input and output structure of the LSTM network includes all the variables $p_o = \{S \quad h \quad \alpha \quad \dot{S} \quad \dot{h} \quad \dot{\alpha}\}^T$ as the output and $p_i = \begin{bmatrix} F_x & F_y & T \end{bmatrix}^T$ as the input with the LSTM input-output mapping defined as follows:

$$[p_i]^{n+1} = f(W[p_o]^n + U h_t^n + b) \tag{4}$$

where the additional matrix $U$ which is absent in the ANN architecture associates input from the previous time steps and introduces the memory effect in the ANN architecture leading to the Recurrent Neural Network (RNN) outlined in Lipton et al. [18]. For the simplest RNN, $h_t$ is the predicted output value at the $n^{th}$ time step that is $\{p_o\}^n$. In the current work, this network is implemented in the *Keras* platform [19]. The input and output structure of such a network is shown in Fig. 2(a) and 2(b) respectively. The input structure is a tensor which contains the first dimension indicating total time steps $n_t$, the second dimension indicating the backpropagation length (memory effect or previous time steps data) in case of input and number of neurons in case of output, and the third one indicating the type of inputs or outputs (i.e., plunge, pitch, forces, plunge velocity, pitch velocity, forces, and moments, etc.).

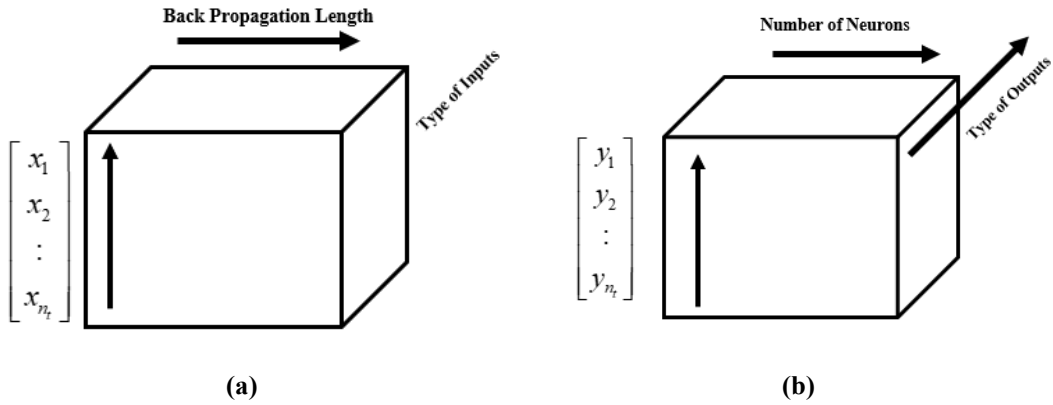

(a) (b)
Figure 2: (a) Input and (b) Output data structure

The limitation of the RNN layer is that it creates instances of vanishing or exploding gradient problems because of the long-term dependencies over the time-sequential dataset (i.e., multiplication of gradient terms over total timestep $n_t$ in a sequence prediction). Hochreiter et al. [20] introduced the LSTM network to mitigate this problem of vanishing or exploding gradients whereby unlike the conventional RNN which keeps its contents of previous time steps while computing the gradients, the LSTM introduces four distinct operations in each RNN cell: memory cell, input gate, forget gate, and output gates. Details of the Memory cell, Input gate, Forget gate and Output gate, and the hidden layer operations in connection with the LSTM network are defined as follows:

- Input Gate: $i^n = \sigma(W_i h^{n-1} + b_i)$
- Forget Gate: $f^n = \sigma(W_f h^{n-1} + b_f)$
- Output Gate: $o^n = \sigma(W_o h^{n-1} + b_o)$ (5)
- Cell State: $c^n = i^n \odot c^{n-1} + i^n \odot \tanh(W_c h^{n-1} + b_c)$
- Hidden Layer: $h^n = o^n \odot \tanh(c^n)$.

where $\odot$ is the Hadamard product, input forces are fed to the LSTM network at each step of the input sequence defined as $n$, i is the Input gate and f is the Forget gate which will choose the necessary information to be passed through or forget through cell state c, $W_f$ and $W_c$ are the corresponding weight matrices, $b_i$, $b_f$ and $b_c$ are the bias vectors for the Input gate, Forget gate and Cell state, respectively. The Output gate, o, on the other hand, decides the control of the flow of the information from the cell state to the next hidden layer. These operations help the LSTM cells to keep only the necessary memories thereby mitigating the occurrence of exploding and vanishing gradient problems in the back-propagation algorithm for each iteration step of the gradient descent optimization. Figure 3(a) shows a schematic of the internal architecture of a single LSTM cell. Figure 3(b) shows the stacking of LSTM cells to enhance the network size. The output dimension of the LSTM layer $\in \mathbb{R}^{n_t \times N}$ where $N$ is the number of neurons, the final layer output must be sent through a dense ANN layer to obtain the required output size which is $\mathbb{R}^{n_t \times 3}$ here ($\{S \quad h \quad \alpha \quad \dot{S} \quad \dot{h} \quad \dot{\alpha}\}^T$. Figure 3(b) also shows the stacked LSTM layers sent to the dense ANN layer to form the final LSTM/LSTM-PINN network.

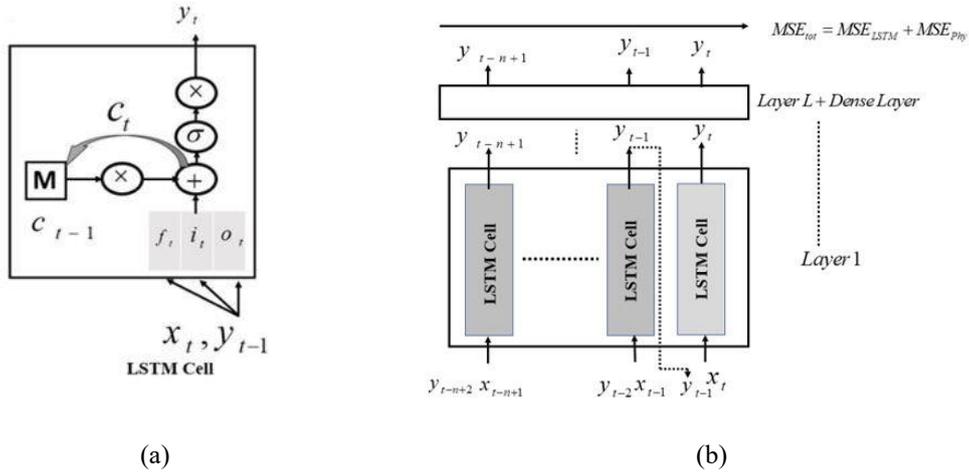

(a)          (b)

**Figure 3:** (a) Structure of a Single LSTM cell (b) Stacked LSTM cells forming LSTM/LSTM-PINN

The loss function associated with the LSTM network, i.e., $MSE_{LSTM}$ is computed using input and output values of the LSTM network. The outputs from the LSTM network are used in the governing equations to estimate the physics informed loss $MSE_{PINN}$. The total loss for a LSTM-PINN network is then expressed as

$$MSE_{LSTM-PINN} = w_1 * MSE_{LSTM} + w_2 * MSE_{PINN} \tag{6}$$

where $MSE_{LSTM} = \frac{1}{n}\sum_{i=1}^{n}(y_{pred} - y_{actual})^2$ and

$$MSE_{PINN} = \frac{1}{n}\sum_{i=1}^{n}\left( \frac{3\begin{Bmatrix}X\\\dot{X}\end{Bmatrix}^{n+1} - 4\begin{Bmatrix}X\\\dot{X}\end{Bmatrix}^{n} + \begin{Bmatrix}X\\\dot{X}\end{Bmatrix}^{n-1}}{2dt} + \begin{bmatrix}0 & 0 & -1 & 0\\ 0 & 0 & 0 & -1\\ M^{-1}K & 0 & 0\\ & 0 & 0 &\end{bmatrix}\begin{Bmatrix}X\\\dot{X}\end{Bmatrix}^{n+1} - \begin{Bmatrix}0\\0\\M^{-1}F\end{Bmatrix}^{n+1}\right)^2 \tag{7}$$

are the losses associated with the LSTM network and the loss associated with the physics of the floating box dynamics respectively. w1 and w2 are the weightage values associated with the data driven and physics-based loss function. The input of the LSTM-PINN network is set as $P_i = [F_x \; F_y \; T]$ whereas the output of the LSTM-PINN network consists of the dynamic responses and their time derivatives, i.e. $\{S \; h \; \alpha \; \dot{S} \; \dot{h} \; \dot{\alpha}\}^T$ of the floating box at the $n_t^{th}$, $n_t^{th}-1$, $n_t^{th}-2$ time steps. Therefore, the size of the input of the network is $\mathbb{R}^{3 \times N_t}$ whereas size of the output of the network is $\mathbb{R}^{18 \times N_t}$. The $n_t$ is any time instant in total time steps of $N_t$.

**II.(b) DEIM-LSTM network**

The manifold hypothesis outlined by Bengio et al. [21] forms the basis of dimensionality reduction which states that the solution of a high dimensional dynamical system lies near a low dimensional manifold $S$ embedded in a large dataset of size $R^n$. The Discrete Empirical Interpolation Method (DEIM) is a data compression method for reconstructing the full data set from the information of the flow variables specified at a few sensor locations as shown in Halder et al. [23]. For example, if the dataset $D$ is a function of time $t$ or any parameter $\mu$, then $D$ can be expressed as $D(t) = \Phi c(t)$ where $\Phi$ defines the DEIM modes and $c(t)$ is the coefficient vector which is computed from an index matrix or mask matrix $P$ used for indexing sensor locations in the computational domain. The index matrix is defined as $P = [e_{\rho 1}, \ldots \ldots, e_{\rho l}]$, $P \in \mathbb{R}^{n \times l}$ where each column is defined as $e_{\rho i} = [0, \ldots 0..1 \ldots 0]^T$ which implies a value of 1 at location $\rho_i$, which is decided from an error minimization-based algorithm as shown in Chaturantabut and Sorensen [9]. Hence, the full dataset can be approximated as:

$$D(t,\mu) = U(P^T U)^{-1} P^T D(t,\mu) \tag{8}$$

where $U(P^T U)^{-1}$ is considered as the DEIM modes $\Phi$ and $P^T D(t)$ indicates the sensor locations values which form coefficient vector $c$. For the large datasets, the CUR approximate matrix decomposition using the DEIM algorithm outlined in Sorensen and Embree [22] can also be used to compute the control points efficiently. The basic construct of the non-intrusive reduced order model (NIROM), results in a dimensional reduction from $R^n$ to $R^m$. The size of an actual dataset is $n$, and the size of the reduced dataset is $m$ where $m \ll n$. This reduced dataset is fed to a deep learning algorithm for exploring the feature dynamics of the. A series of snapshots, each having the size of $R^n$ is reduced to a set of snapshots of the size of $R^m$ before feeding into the deep-learning network.

For the prediction of the wave-structure interaction, Discrete Empirical Interpolation Method (DEIM) is coupled with the LSTM network for the floating box problem with three degrees of freedom (pitch, plunge ad surge), and the reader is referred to details of this step in Halder et al. [23]. Figure 4 shows the schematic of the application of the DEIM-LSTM to predict the dynamics of a floating box in an unsteady hydrodynamic wave. First, the unsteady wave elevation distribution data is generated from the *DualSPHysics* simulation by exciting the wavemaker so that the computed data of hydrodynamic loads on the box and the kinematics of the floating box forms the ground truth or the actual data. The DEIM algorithm is then applied to the spatio-temporal wave height distribution and a few control points are selected for determining the DEIM modes. For the training phase of the LSTM/LSTM-PINN network, LSTM is applied to the wave heights at these control points and the predicted box motion from actual data. After the LSTM network has been trained sufficiently, it can be used to predict the box motion under any arbitrary wavemaker excitation profiles.

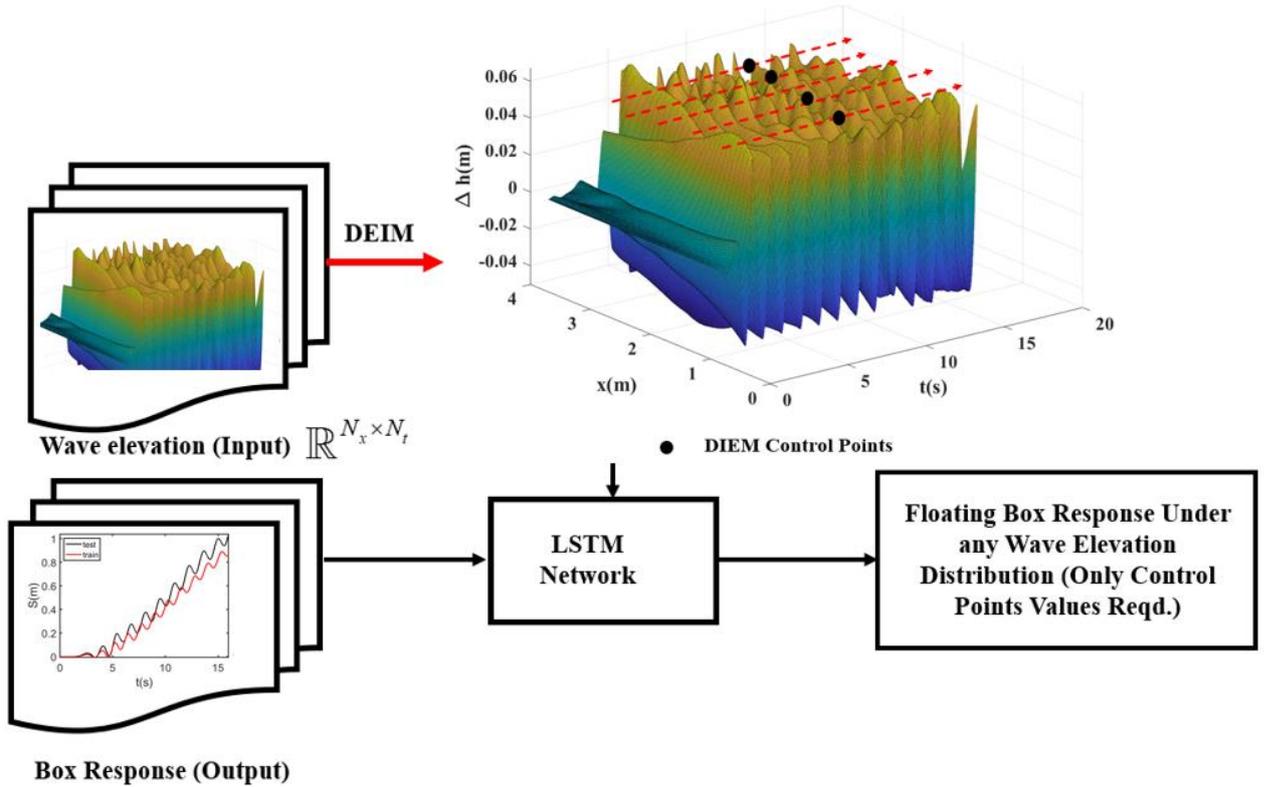

**Figure 4:** Schematic of the DEIM-LSTM Network.

## III. Results and Discussion:

In this section, the prediction of the floating box dynamics due to the surface wave elevations in the wave basin is discussed. First, the DEIM-LSTM is used for predicting the dynamics of the floating box where the wave elevation is the input. In the next section, the governing equations for the dynamics of the floating box are coupled with the data-driven loss function in the LSTM network.

Figure 5 compares the temporal dynamical motion of the floating box responses of computed surge, heave and pitching with the experimental data of Ren et al. [16] corresponding to the paddle in the wavemaker in the wave-basin being excited with a sinusoidal function of period 1.2 s and amplitude of 0.1 m to validate the SPH model predictions with experimental data. This is the ground truth for the current study.

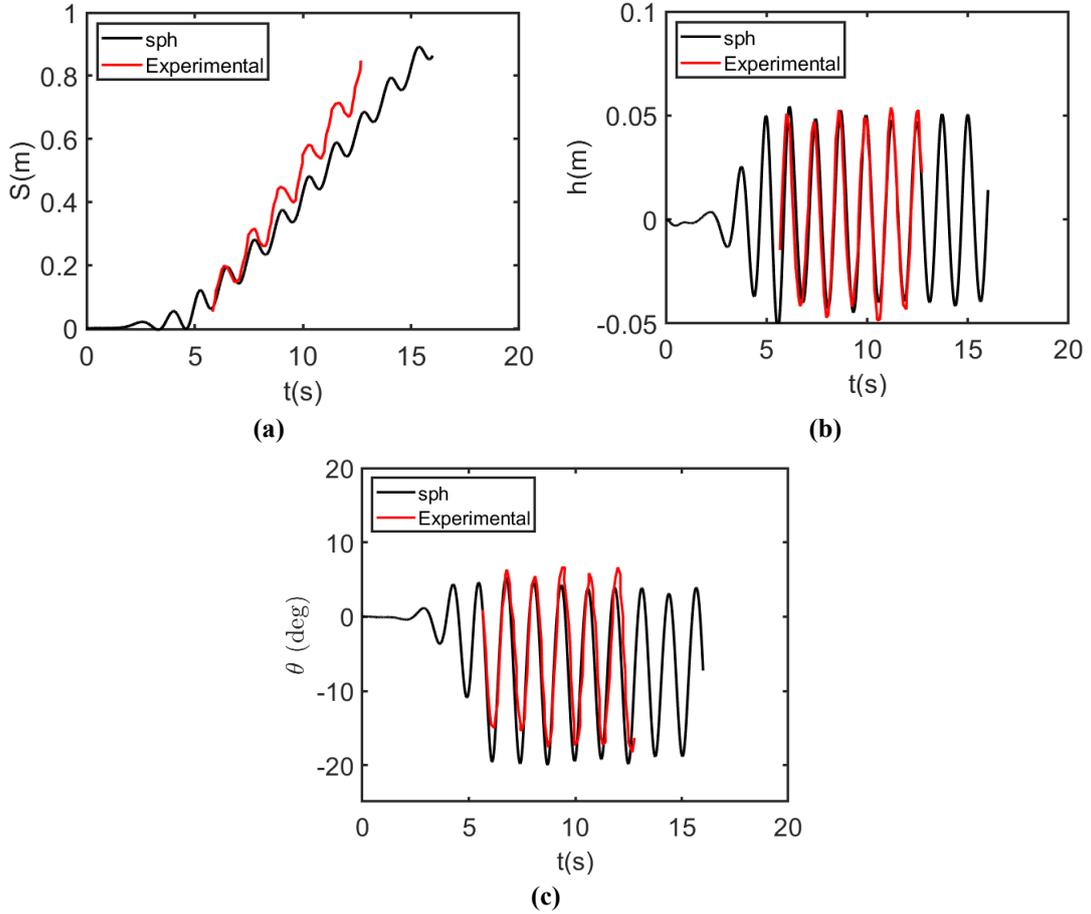

**Figure 5:** Comparison of the box response (a) surge (b) heave and (c) pitch motion using experiments and current numerical model

### III. a DEIM-LSTM Prediction of Wave-Floating Box Interaction Dynamics

The paddle in the wavemaker is excited with a sinusoidal function of period 1.2 s and amplitude of 0.1 m for the training of the DEIM-LSTM network and a sinusoidal function of period 1.3 s and amplitude of 0.12 m as a test function. The temporal-spatial distribution of the wave elevations corresponding to these scenarios is shown in Figures 6(a) and 6(b).

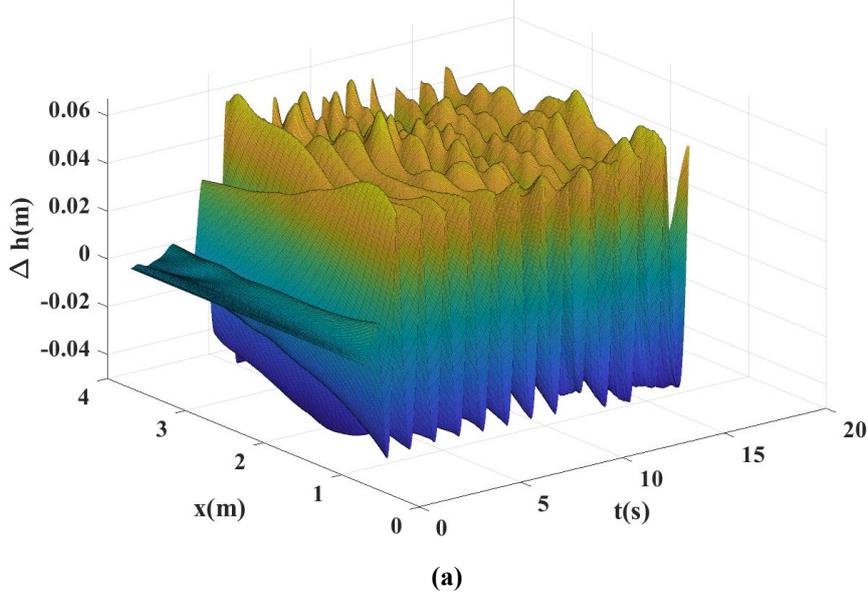

(a)

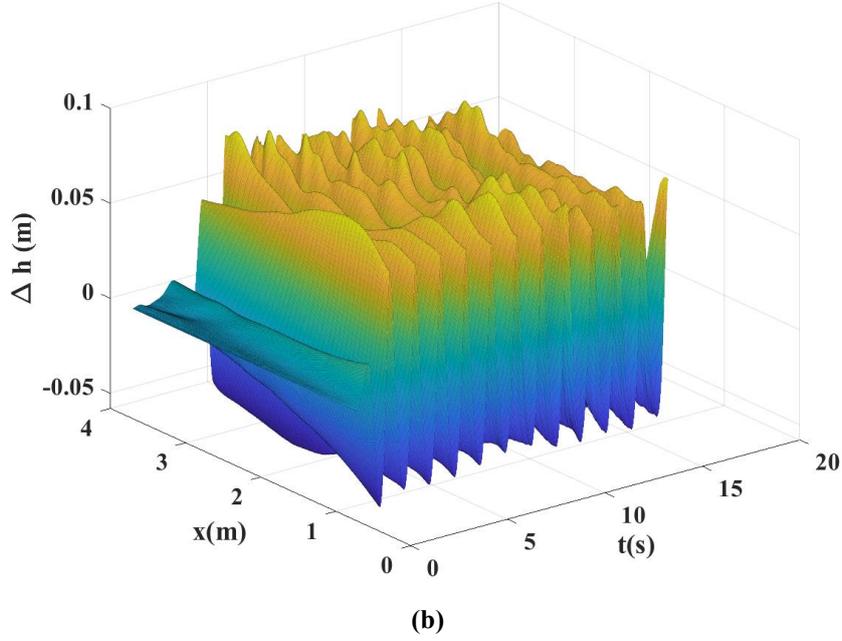

(b)

**Figure 6:** (a) Training and (b) Test elevation of the wave surface considered as an input to the LSTM network.

The training and test wave elevation data is of size $\mathbb{R}^{N_x \times N_t}$ where $N_x$ is the 300 spatial locations and $N_t$ is the 2001 temporal locations for this study. DEIM is then applied to dimensionally reduce this large input dataset of Fig. 6 from $\mathbb{R}^{N_x \times N_t}$ to $\mathbb{R}^{m_x \times N_t}$ ($m_x \ll N_x$). The Deep Learning-based approach is considered at this point to learn the temporal dynamics. Since the LSTM approach is already a computationally efficient approach as compared to the SPH model, the dimensionality reduction is considered only in the spatial direction. The Singular Value Decomposition (SVD) is first applied to the input dataset to decide on an optimal number of modes to represent the temporal-spatial distribution of the wave elevation considered as an input to the LSTM network as shown in Fig.7. Any dynamical system can be represented as a combination of few modes and the singular values are associated with the energy contained by those modes. Figure 6 shows the variation of the scaled singular values (scaled with the largest singular value, $Sv_{max}$) which indicate the number of modes that contain most of the energy of the dynamic system and after a few modes, the singular values drop to a very small value which is not important for system dynamics.

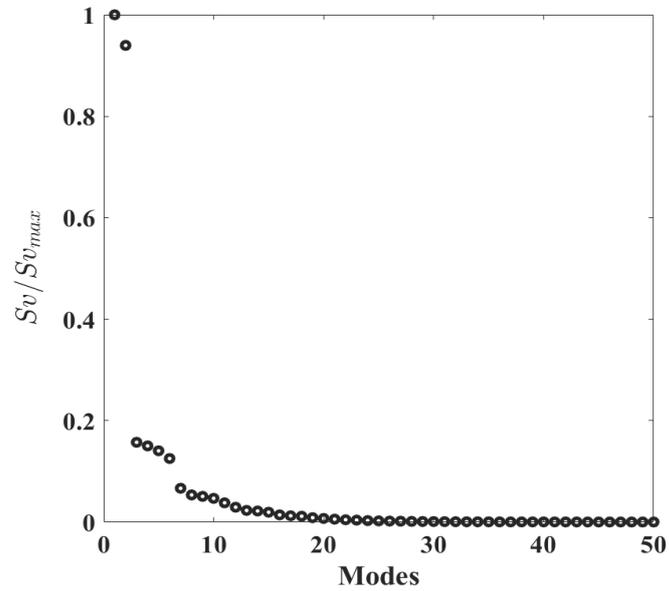

**Figure 7:** Normalized Singular Values after the application of the SVD on the input data.

Figure 8 shows a selection of DEIM control points for the reconstruction of instantaneous wave profile. Among the 300 spatial points, only 30 points indicated as $x_1, x_2, ....x_n$ are chosen for all the temporals.

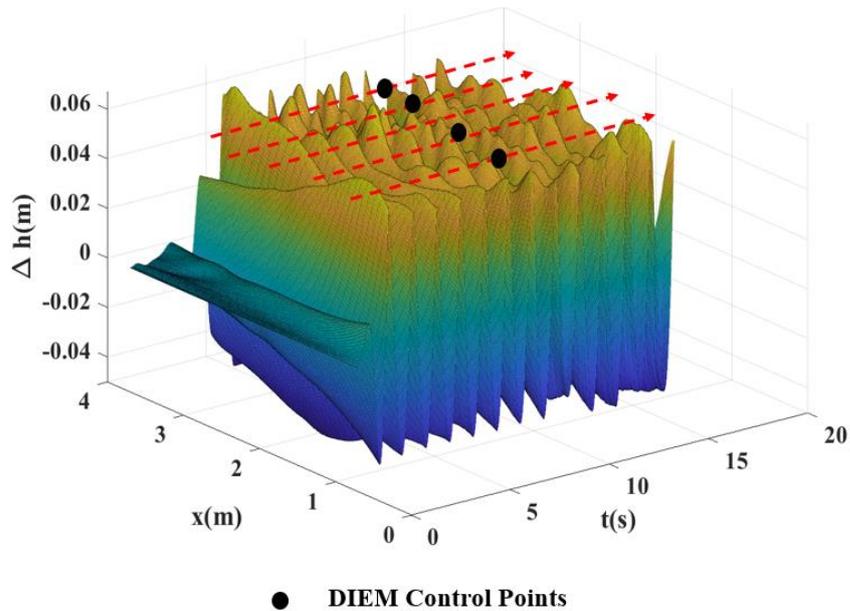

**Figure 8:** DEIM Control points marked as black dots at a time instant.

locations to significantly reduce the size of the input data to $\mathbb{R}^{30 \times 2000}$ noting that $\Delta h$ is the difference between the instantaneous wave elevation and initial wave elevation. The points $x_1, x_2, \ldots x_n$ are termed as DEIM control points and the DEIM modes termed as $\Phi$ in Section II. B contains the inherent features of the wave dynamics. Among the 30 DEIM modes corresponding to these control points, the spatial variation of the first 4 modes is shown in Fig.9(a). If the values of the DEIM control points for the training and test wave distribution as shown in Fig.8 are known, then the entire spatio-temporal distribution of the wave can be reconstructed as shown in Fig. 9(b). The zoomed view of region marked by the red box in Fig.9(b) is shown in Fig.9(c) to showcase the accuracy of the prediction of the test wave elevation at the 12 s instant by comparing the predictions using different numbers of DEIM modes ranging from 10 to 30 in steps of 10 with the computed ground truth or actual wave height (computed using SPH) illustrating the adequacy of the selected number of DEIM control points. It shows that if 10 DEIM modes are selected the reconstructed signal deviates from the actual wave height whereas, after 20 modes, the change in the reconstructed wave height is insignificant. Figure 10(a) to 10(c) show that the error distribution drastically drops when the number of selected modes is increased from 10 to 30.

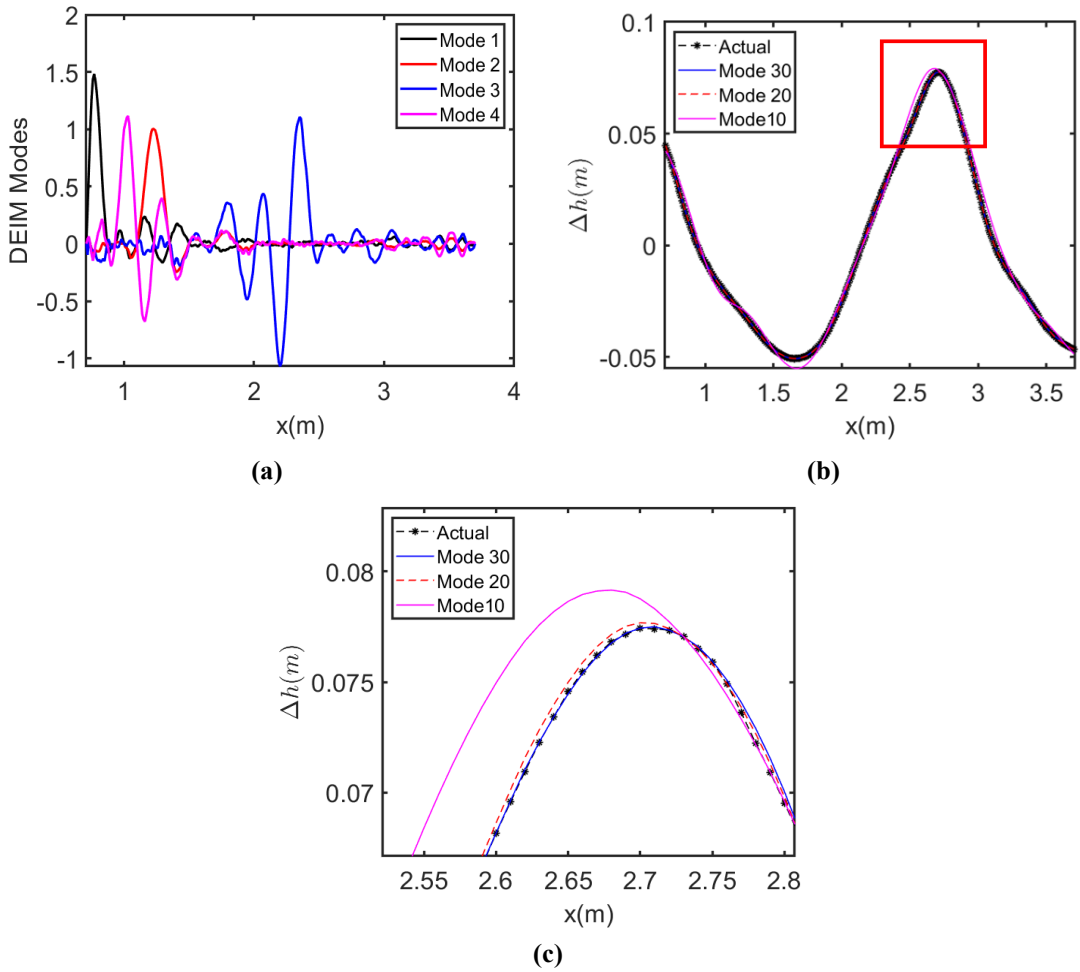

**Figure 9:** (a) First four DEIM Modes and (b) comparison of predicted wave elevation using different numbers of the DEIM modes (c) zoomed view of the red box in (b) compared to computed wave elevation at the 12 s instant (ground truth or actual data).

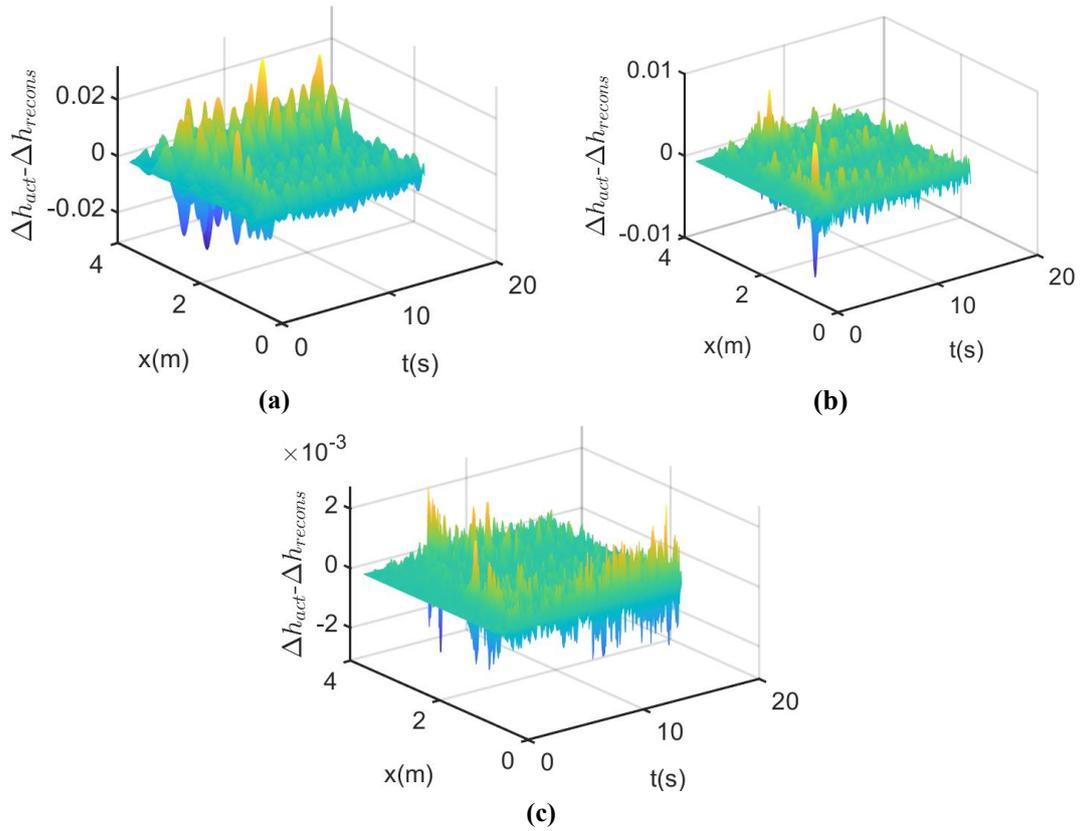

**Figure 10:** Reconstruction error distribution of the test elevation using (a) 10 (b) 20 and (c) 30 DEIM modes of the training wave elevation.

Figures 11 (a)-(c) compares the prediction of temporal variation of heave, surge, and pitch angle dynamics of the floating box respectively for both training and test wave elevation distributions shown respectively in Fig. 6(a) and 6(b) and which will be used for DEIM-LSTM predictions.

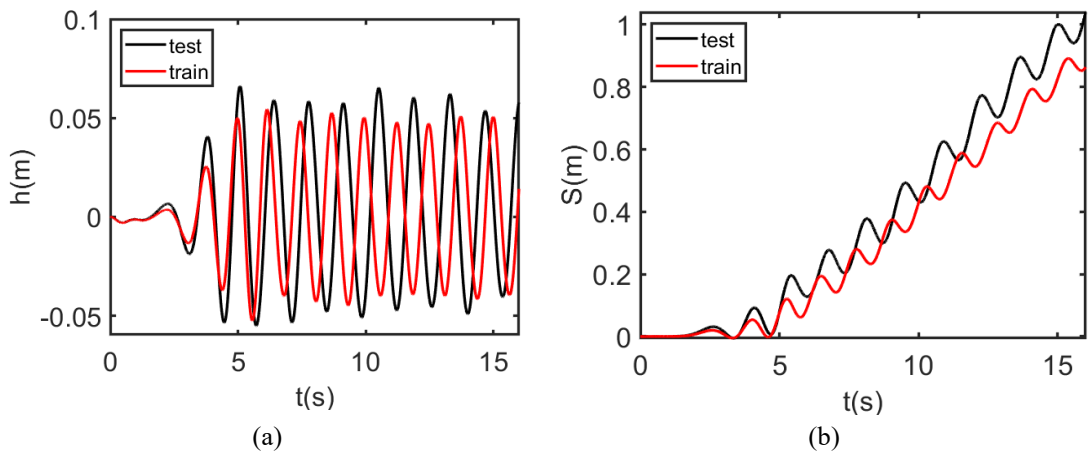

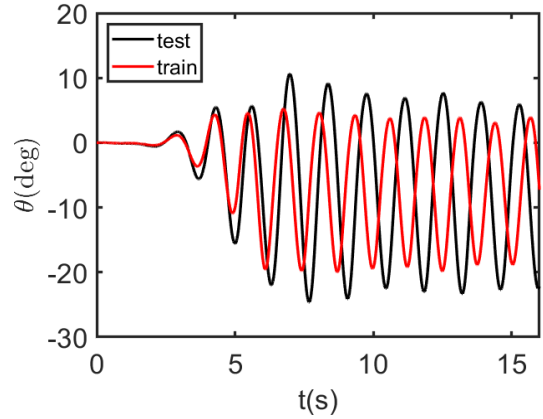

(c)

**Figure 11:** Temporal variation of (a) heave (b) surge and (c) pitch dynamics of the floating box computed using SPH used as training and test data for DEIM-LSTM predictions.

A LSTM network consisting of 25 input time sequence termed as *n* in Fig.3 of the Section II.a, 4 hidden layers, 10 neurons in each layer, and *tanh* activation function is considered for prediction. Figure 12(a)-(c) compares the temporal variation of floating box dynamics of heave, surge, and pitching predicted using DEIM-LSTM with actual data computed from SPH when the training and test wave elevations and the corresponding box motions are the same. Since the training and test dataset are the same, the predicted results have matched significantly well with the high-fidelity simulation results.

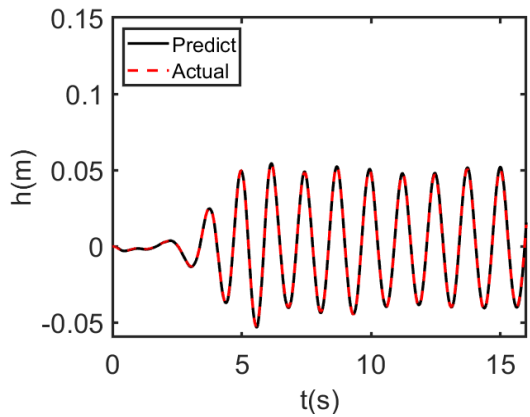

(a)

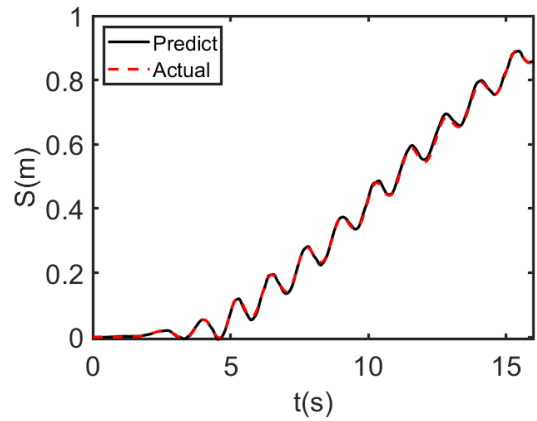

(b)

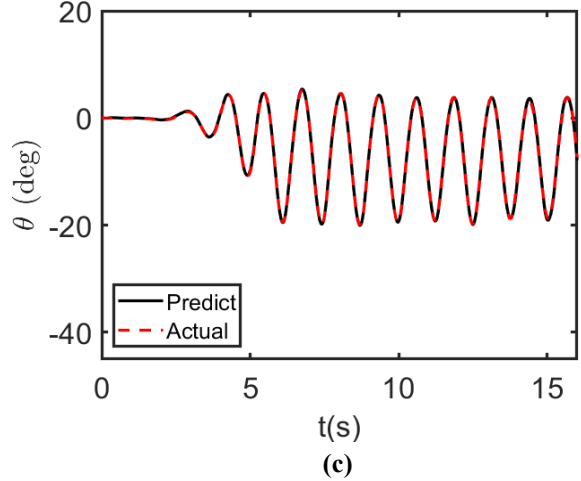

**Figure 12:** Comparison of the temporal variation of (a) heave (b)surge (c) pitch response of floating box from DEIM-LSTM prediction with SPH predictions for the case when the training and test dataset are the same.

Next the prediction of the floating box dynamics using DEIM-LSTM when the training and test datasets used are different is considered and the impact of some of the pertinent network hyperparameters, namely the number of epochs reached during training, the size of the backpropagation length (termed here as input sequence) and the number of neurons in the hidden layers on the prediction is assessed. Figures 13(a)-(b) compares the DEIM-LSTM prediction of the heave, surge, and pitch motion of the floating box with the actual data from SPH simulation for number of epochs ranging from 150-250-500 for which the training time ranges from 344s, 599 s and 1264s respectively. Mean Absolute Errors (MAE) associated with the surge, heave, and pitch response corresponding to epoch 150 are 0.0251, 0.0164, and 8.29 respectively whereas for 250 and 500 epochs they are 0.0241, 0.0152, 8.54, and 0.0218, 0.0153 and 8.3089 respectively. Although the training time increases almost linearly with the number of epochs, the prediction accuracy remains the same and compares well with the actual data. Figure 13(d) shows the mean squared error variation with the number of epochs. It shows that the training error does not further decay after 300 epochs.

The effect of varying the size of the backpropagation length or the training input sequence on the DEIM-LSTM prediction capability is shown in Fig. 14 for the same data sets. For the input sequence of 15, 25, and 35, the training time is 378s, 599s, and 775s respectively. For all three cases, the predicted result matches well with the actual data. The MAE associated with the input sequence of 15 and 35 are 0.0224, 0.0145, 7.808, and 0.0248, 0.0171, 8.26 respectively for the surge, heave and pitch motion while the MAE of the input sequence 25 is indicated earlier for the epoch variation of 250. It shows that the prediction accuracy is better at the lower input sequence.

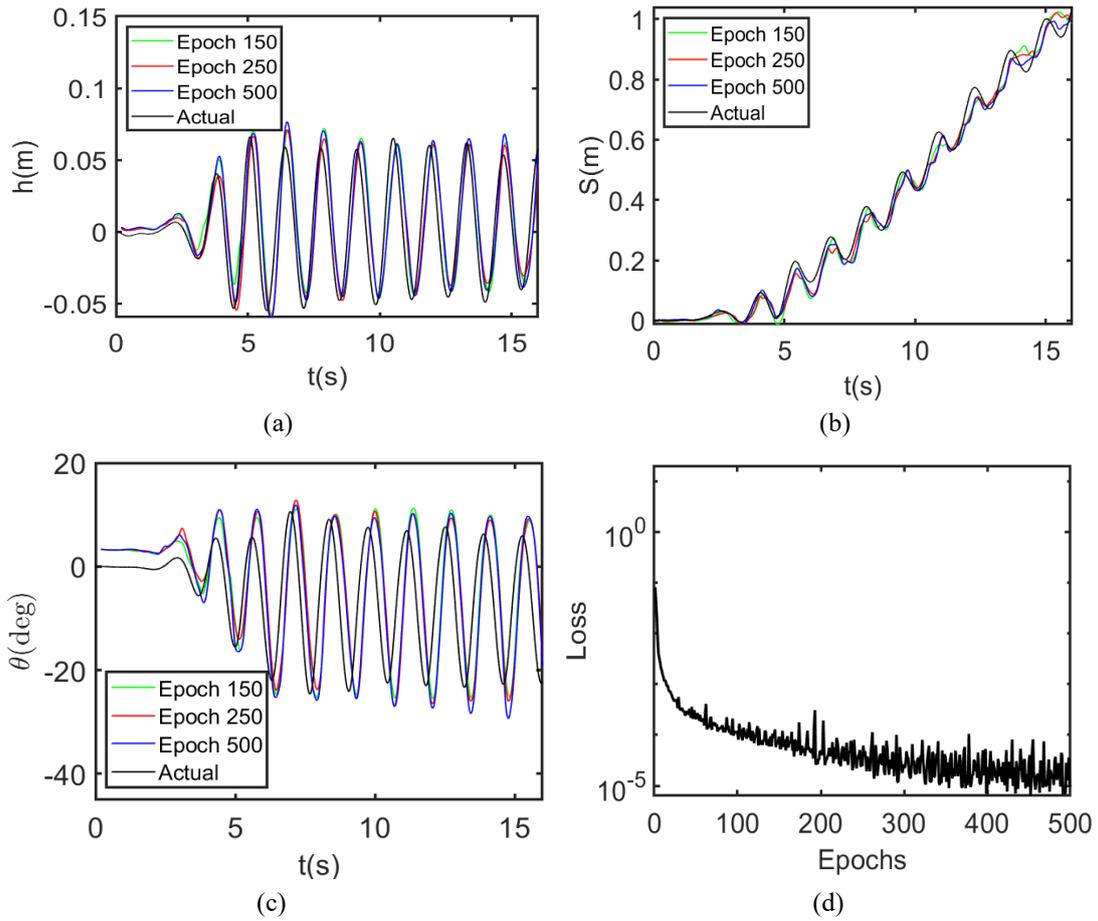

**Figure 13:** Comparison of the temporal variation of (a) heave (b) surge (c) pitch response of floating box from DEIM-LSTM prediction with SPH predictions for the case when the training and test dataset are different for different epoch length. (d) Variation of the mean squared loss history.

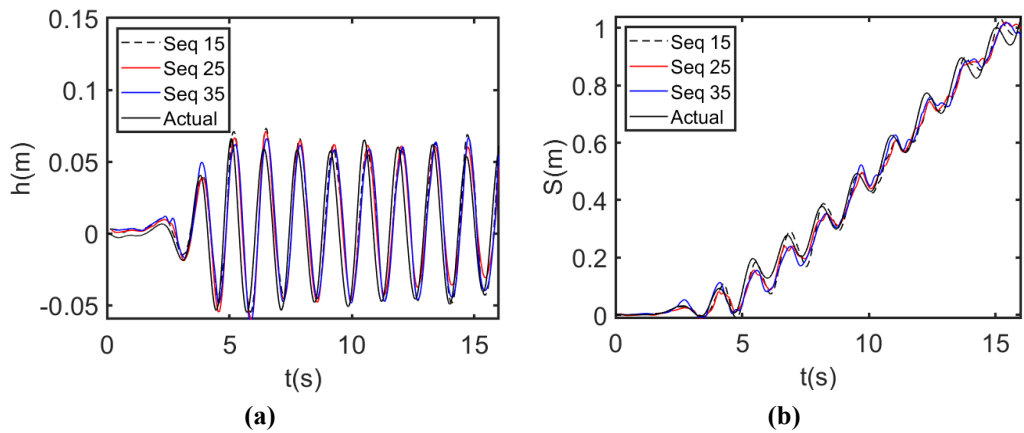

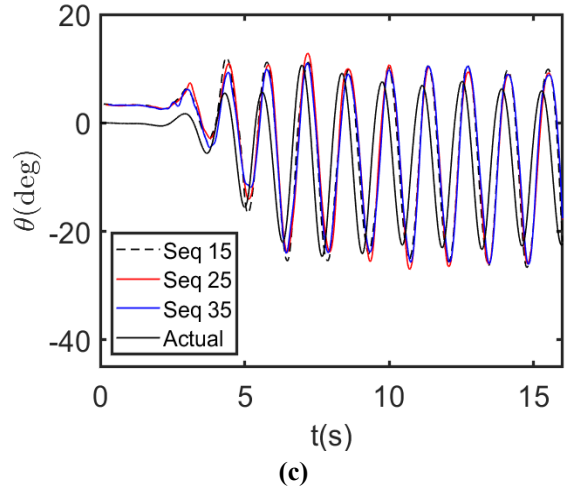

**Figure 14:** Comparison of the temporal variation of (a) heave (b)surge (c) pitch response of floating box from DEIM-LSTM prediction with SPH predictions for the case when the training and test dataset are different for different input sequence.

The effect of varying the number of neurons in the hidden layers on the prediction capability of the DEIM-LSTM is shown in Fig. 15(a)-(c) for the same data sets.

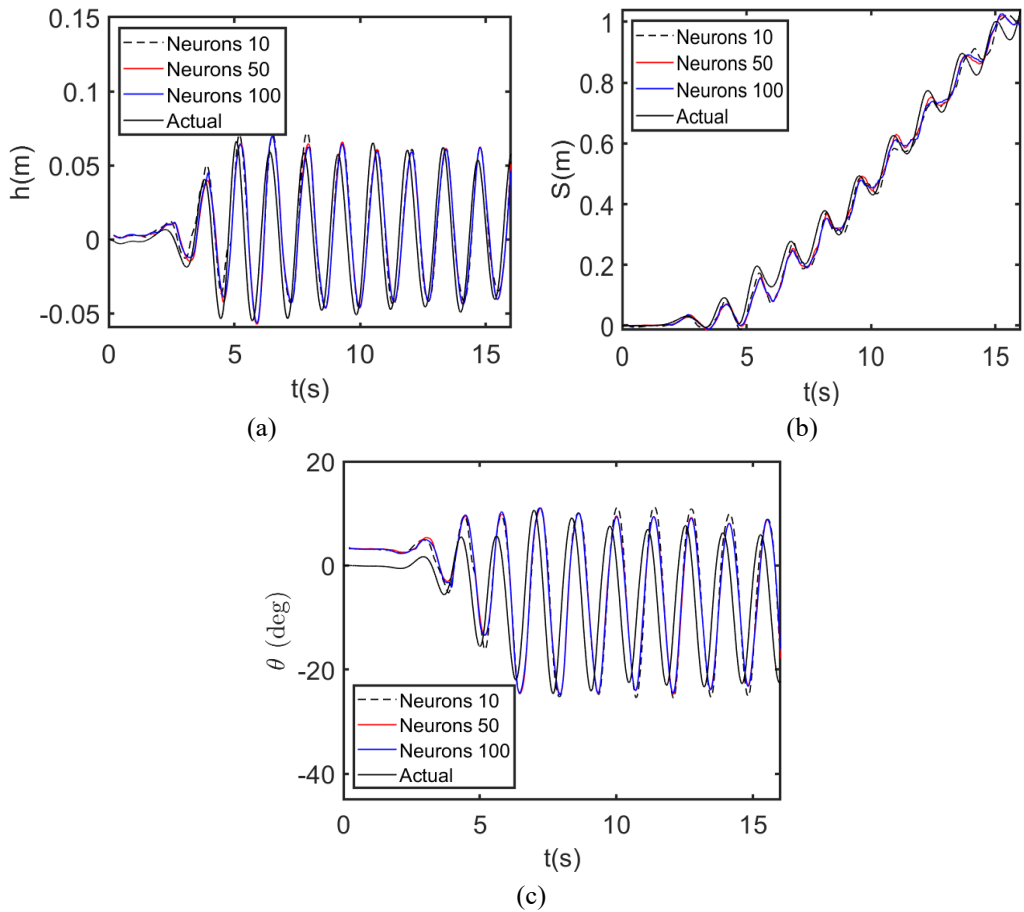

**Figure 15:** Comparison of the temporal variation of (a) heave (b)surge (c) pitch response of floating box from DEIM-LSTM prediction with SPH predictions for the case when the training and test dataset are different for different input sequence.

As the number of neurons in each layer is varied from 10 to 50 to 100, and the training time vary from times 599s, 583s to 1637s. The training time is not linearly scaled like the previous two cases of epoch numbers and input sequence variation. The MAE error associated with the 50 neurons and 100 neurons are 0.0223, 0.0169 8.0429 and 0.0240, 0.0167 7.95 respectively for the surge, heave, and pitch motion whereas the MAE error of 10 neurons is indicated earlier for the epoch variation of 250 and input sequence of 25. The deviation of the predicted pitch response from the actual data decreases with the increase in the number of neurons.

For all three variations such as the number of epochs, input sequence, and the number of neurons, the pitch response shows a constant phase shift from the benchmark result which is also evident from the MAE associated with the pitch response as well. If the neural network is trained with the input wave elevation and corresponding box motion generated by a multi-harmonic paddle excitation instead of a single frequency signal, it may be able to capture the inherent nonlinearity in the flow physics and this deviation of the predicted result from the actual data is expected to get minimized.

**III.b LSTM-PINN Prediction of Wave-Floating Box Interaction Dynamics:**

In this section the rigid body dynamics i.e., Eqn. 1 is incorporated to enhance the LSTM network loss function to construct a physics informed network (LSTM-PINN) and to assess the effect of training data size on the quality of the predictions. The SPH computed hydrodynamic forces are applied to the box motion which are considered as an input of the LSTM network and box motion i.e., pitch, plunge, and surge motion are taken as output. The high frequency noise content in the surge, heave and pitch responses of the box motion is first smoothed out and the acceleration signals based on the smoothed displacements are computed using the second-order backward Euler method. The noisy dataset generates spurious residual in the optimization process of the learning algorithm which often gets stuck at a local minimum and thereby results in poor prediction of the overall LSTM-PINN method. Figure 16(a)-(c) compares the smoothed accelerations in the surge, heave, and pitch axes with the actual computed accelerations from SPH and it shows that even if the displacement signal is smoothened by removing the high-frequency numerical noise, obtained acceleration matches the mean value of the actual acceleration. After the initial transients die out (~ 5s), the actual acceleration coincides with the smoothened signal. As the aim of the present effort is to show a proof of concept of the proposed surrogates, a force signal based on the smoothed displacement is used instead of the actual force obtained from the SPH simulation. Both the training and the test dataset for this network are the same as that shown in Fig. 6(a) and the training box responses in Fig.11. A percentage of the training input and output is considered as the training dataset of the network, and the aim is to predict the entire output response using the partial data used for training.

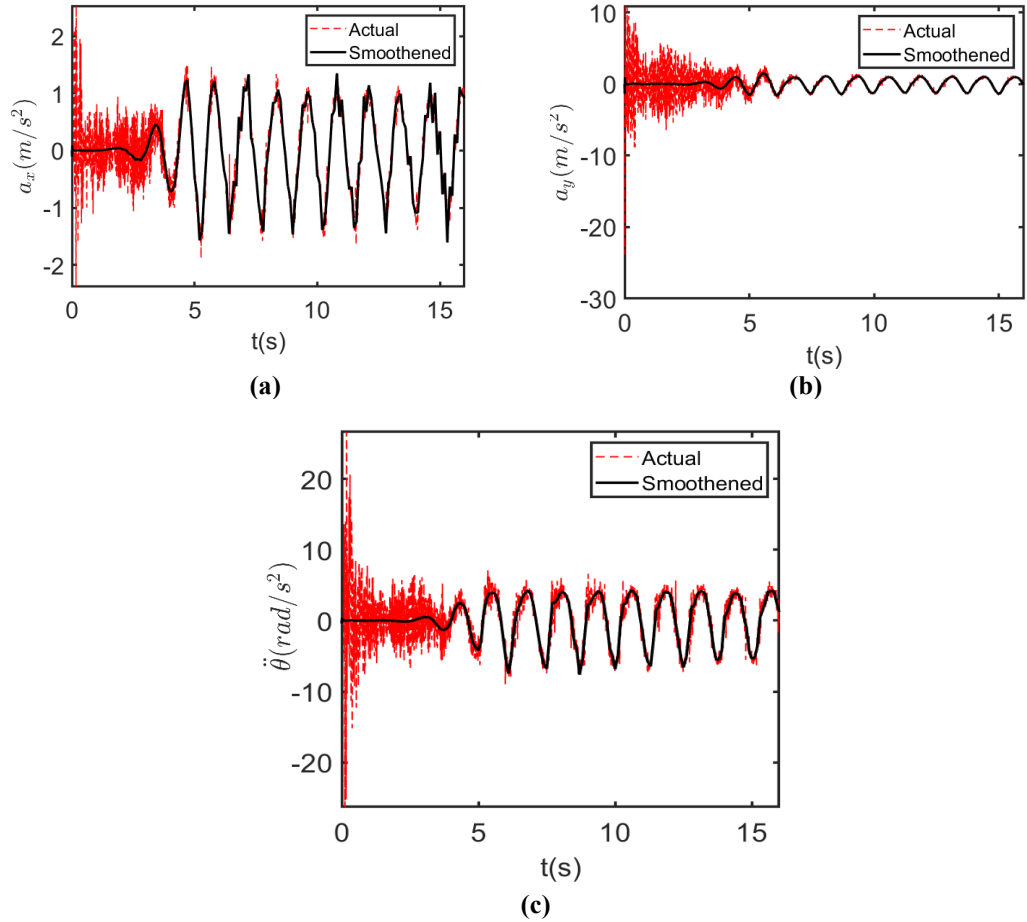

**Figure 16:** Smoothening of the acceleration signal in (a) heave (b) surge (c) pitch direction

The implication of using LSTM-PINN is that the size of the training data can be reduced since the incorporation of the physics will correct the weights and biases accordingly. Figure 17(a)-(c) and Fig.18(a)-(c) show the LSTM and LSTM PINN prediction of heave, surge and pitch responses using 59.87 % and 0.2% data. For both the cases, the heave and pitch motions are predicted well using the LSTM-PINN network whereas the predicted surge motion using 0.2% data does not fit well with the benchmark result.

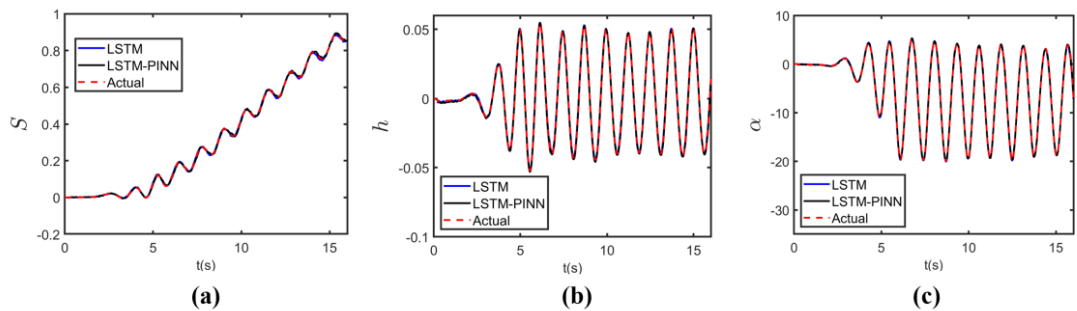

**Fig 17:** Comparison of DEIM-LSTM and DEIM-LSTM-PINN (trained using 59.87% of the data prediction of the temporal variation of (a) heave (b) surge (c) pitch response of floating box with actual data.

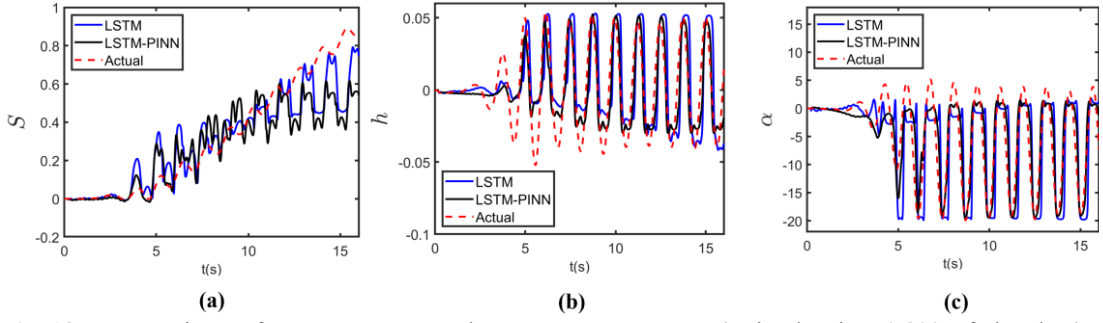

**Fig 18:** Comparison of DEIM-LSTM and DEIM-LSTM-PINN (trained using 0.2% of the data) prediction of the temporal variation of (a) heave (b) surge (c) pitch response of floating box with actual data computed using SPH.

Table 1 compares the L2 error of the prediction accuracy using DEIM-LSTM and DEIM-LSTM-PINN based on the percentage of the training data (ground truth or actual data computed using SPH) used. Table 1 shows that for sparse data (~0.2%) the LSTM-PINN offers a better prediction of pitch and heave motion. If the training data used in large, LSTM predictions appear to be better than LSTM-PINN thereby suggesting that with the use of big data physics may not have significant influence on the prediction.

| | **Table 1:** L2 Error in percentage for Surge, heave, and pitch motion for DEIM-LSTM and DEIM-LSTM-PINN prediction as compared to the benchmark data. | | |
|---|---|---|---|
| | **Percentage of Data Used** | **LSTM (% L2 error)** | **LSTM-PINN (% L2 error)** |
| | ~100% | 0.0013 | 0.0085 |
| | ~59.87% | 0.0048 | 0.0098 |
| $S$ | ~0.20% | 8.661 | 13.4368 |
| | ~100% | 0.0060 | 0.0233 |
| | ~59.87% | 0.0293 | 0.0654 |
| $h$ | ~0.20% | 57.67 | 24.81 |
| | ~100% | 0.0044 | 0.0051 |
| | ~59.87% | 0.0103 | 0.0214 |
| $\alpha$ | ~0.20% | 42.2455 | 12.86 |

From Table 1 the LSTM network outperforms the LSTM-PINN network despite the introduction of physics loss in the total loss function. Figure 19 compares the variation of the mean squared error-based loss function as shown in Eqn.6. It appears that the loss associated with the LSTM prediction decays faster than that of the LSTM-PINN prediction when the complete dataset is used. The additional physics-based loss function increases the residual values, and if the parameters of the optimization algorithms are not set properly the residual values often get stuck in a local minimum point thereby deteriorating the prediction accuracy as compared to the LSTM algorithm. Therefore, different weightages (w1 and w2 in Eqn.6) of the loss functions are considered and their effect on the decay in mean squared error (MSE) are compared in Fig.20. The weightage matrix of the loss function w2 are as follows:

$$\begin{aligned} w2a &= [1 \quad 1 \quad 1 \quad 1 \quad 1 \quad 1], \\ w2b &= [1 \quad 1 \quad 1 \quad 0.001 \quad 0.001 \quad 0.001], \\ w2c &= [0.01 \quad 0.01 \quad 0.01 \quad 0.001 \quad 0.001 \quad 0.001], \end{aligned} \quad (9)$$

*w1* and w2 are associated with the data-based loss function $MSE_{LSTM}$ and $MSE_{PINN}$ respectively. The LSTM-PINN algorithms show better accuracy only for the *w2c* which also indicates the minimization of the residuals and the *MSE* values are decaying faster than LSTM approach. With the *w2a* and *w2b* weightages the loss functions do not decay beyond a certain value thereby resulting in a very poor prediction of the box responses even though the governing equations-based loss functions are considered.

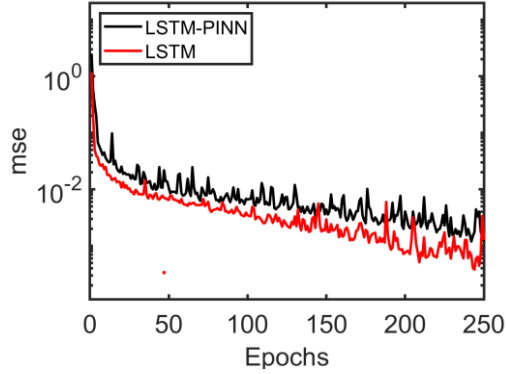

**Figure 19:** Loss history for full dataset with LSTM and LSTM-PINN algorithm

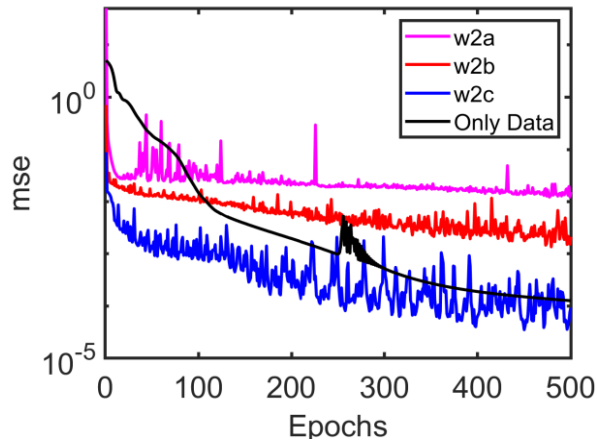

**Figure 20:** Decay of mese values with different weightage of the loss function.

**IV. Conclusion:**

A novel non-intrusive reduced order model is introduced for the prediction of the temporal variation of the surge, heave and pitching dynamics of a freely floating box on the wavy surface of a water wave maker basin. A dimensionality reduction approach DEIM is coupled with a LSTM-based deep learning approach for the reduction of the computational training time. The neural network parameters are varied to assess their influence on the prediction accuracy. It is noticed that there is a constant phase shift of the predicted pitch motion from the benchmark result, and it is expected that introduction of a mixed harmonic paddle excitation will learn the inherent nonlinearity present in the hydrodynamics and the prediction accuracy can be improved. Finally, the governing rigid body dynamics equation is coupled with the LSTM network in the discrete form to create the LSTM-PINN network. The addition of the physics-based loss function increases the magnitude of the residual of the LSTM network, resulting in a tendency for the loos function to get stuck at a local minimum frequently if the optimization algorithm in the network is not tuned. In the current work, the LSTM-PINN network is applied to the simple structural dynamical system which can be very well learned by the LSTM network. The introduction of the physics-based governing equation in the loss function did not make any significant improvement in the prediction accuracy. On, the other hand, the use of a larger dataset, and the introduction of the physics-based loss function appear to deteriorate the performance by increasing the magnitude of the residual. The potential of the LSTM-PINN will be realized if the LSTM and LSTM-PINN network is also be applied to the SPH equations to form an LSTM/LSTM-PINN network mapping between spatiotemporal coordinates and unsteady flow variables such as velocities and pressure which the authors will address and also explore the coupling of the networks for fluid flow and structural dynamics as a possible set of digital twins for box dynamics in their future work.

**Acknowledgment:** The authors acknowledge the efforts of Ouyang Zhenyu and Tang Xiaoyang for their prompt generation of computational data sets using *DualSPHysics* which serves as the ground truth (actual) data and training data for this work.